\documentclass[a4paper,11pt]{article}
\pdfoutput=1 
\usepackage{grffile}
\usepackage{aas_macros,braket}
\usepackage{jcappub,natbib} 
\usepackage[T1]{fontenc} 

\title{Measuring chiral gravitational waves in Chern-Simons gravity with CMB bispectra}

\author[a,b,c]{Nicola Bartolo,}
\author[a,b]{Giorgio Orlando,}
\author[d]{and Maresuke Shiraishi}

\affiliation[a]{Dipartimento di Fisica e Astronomia ``G. Galilei'', Universit\`a degli Studi di Padova, via Marzolo 8, I-35131, Padova, Italy}
\affiliation[b]{INFN, Sezione di Padova, via Marzolo 8, I-35131, Padova, Italy}
\affiliation[c]{INAF-Osservatorio Astronomico di Padova, Vicolo dell'OSservatorio 5, I-35122 Padova, Italy}
\affiliation[d]{Department of General Education, National Institute of Technology, Kagawa College, 355 Chokushi-cho, Takamatsu, Kagawa 761-8058, Japan}

\emailAdd{nicola.bartolo@pd.infn.it}
\emailAdd{giorgio.orlando@phd.unipd.it}
\emailAdd{shiraishi-m@t.kagawa-nct.ac.jp}

\abstract{Chern-Simons gravity coupled to the scalar sector through a generic coupling function $f(\phi)$ can be tested at the very high energies of the inflationary period. In 
ref.~\cite{Bartolo:2017szm}, we computed the theoretical parity breaking signatures of the $\Braket{ \gamma \gamma \zeta }$ primordial bispectrum which mixes two gravitons and one scalar curvature perturbation. We defined a parameter $\Pi$ which measures the level of parity breaking of the corresponding bispectrum. In this work we forecast the expected $1 \sigma$ error on $\Pi$ using the cosmic microwave background (CMB) angular bispectra. We find that, given the angular resolution of an experiment like {\it Planck}, $\Pi \sim 10^6$ is detectable via the measurement of $BBT$ or $BBE$ angular bispectra if the tensor-to-scalar ratio $r = 0.01$. We also show that, from the theoretical point of view, $\Pi$ can be greater than $10^6$. Thus, our conclusion is that $BBT$ or $BBE$ CMB angular bispectra can become an essential observable for testing Chern-Simons gravity in the primordial universe.}

\begin{document}



\maketitle
\flushbottom

\section{Introduction}
In ref.~\cite{Bartolo:2017szm} we analyzed the theoretical bispectra of primordial perturbations arising from a Chern-Simons modified gravity term coupled to the inflaton field through a generic coupling function $f(\phi)$. In particular, the action of the model reads
\begin{equation} \label{eq:L_studied}
S =\int d^4x \, \left[ \sqrt{g} \left( \frac{1}{2}M_{Pl}^2 R-\frac{1}{2}g^{\mu\nu}\partial_\mu\phi\partial_\nu\phi-V(\phi) \right) + f(\phi)\epsilon^{\mu\nu\rho\sigma}C_{\mu\nu}{}^{\kappa\lambda}C_{\rho\sigma\kappa\lambda} \right]  ,
\end{equation} 
where $g= -det [g_{\mu \nu}]$, $M^2_{Pl}= (8 \pi G)^{-1}$ is the reduced Planck mass and $C_{\mu \nu \rho \sigma}$ is the so-called Weyl tensor, i.e. the traceless part of the Riemann tensor $R_{\mu \nu \rho \sigma}$. In this action the Chern-Simons term is given by the last contribution, i.e. the contraction of the Weyl tensor with its dual. Due to the presence of one Levi-Civita pseudo-tensor, this term breaks both parity and time reversal symmetries of Einstein gravity, preserving the CPT symmetry. The Chern-Simons term arises naturally when we consider a more general theory of gravity beyond Einstein General Relativity, given by a series expansion of covariant terms that contain an increasing number of derivatives of the metric tensor in the action (see e.g. ref.~\cite{Weinberg:2008}).  In particular, the Chern-Simons operator contains four derivatives of the metric, and is the parity breaking term with the lowest number of derivatives in this expansion (for an example of parity breaking operators with higher derivatives see refs.~\cite{Maldacena:2011,Shiraishi:2011st,Soda:2011}). The Chern-Simons term has also some interesting peculiarities. First of all, this term vanishes if computed on the Friedmann-Robertson-Walker (FRW) background metric (the latter is conformally flat and thus the Weyl tensor is zero in FRW). As a consequence, the Chern-Simons term does not modify the background dynamics of inflation and any effect arises only at the perturbation level. This is particularly interesting if we consider that any parity breaking effect might be left imprinted in the primordial perturbations at very high energies, thus in principle we could find signatures of this kind of gravity in CMB anisotropies. Moreover, the Chern-Simons term is a total derivative term, thus the coupling with the inflaton field $f(\phi)$ during inflation is necessary to write a non-trivial theory. Therefore Chern-Simons gravity could have left signatures only on primordial perturbations since at the end of the inflationary epoch the inflaton decays and the Chern-Simons term becomes a surface term, restoring standard gravity.

In the literature effects of Chern-Simons gravity on gravitational waves in an inflationary context have been studied in several works (see e.g. refs.~\cite{Lue:1999,Jackiw:2003, Alexander:2005, Alexander:2006, Alexander:2007, Satoh:2008,Alexander:2009,Satoh:2010, Malenknejad:2012, Myung:2014, Alexander:2016,Baumann:2016, Kawai:2017, Cai:2017, Afkhami-Jeddi:2018own}). One of the most intriguing predictions of the model is that primordial gravitational waves acquire chirality in the power spectrum statistics or, better to say, Right (R) and Left (L)-handed  primordial gravitational waves get a different dynamical evolution during inflation, and this is the first immediate consequence of the breaking of parity symmetry. However, as showed e.g. in ref.~\cite{Satoh:2010}, the amount of chirality produced during inflation has to be very small to prevent the formation of instabilities during inflation. Moreover, in ref.~\cite{Gerbino:2016mqb} it was shown that the CMB angular power spectra are not able to provide a useful constraint on very small chirality. This motivated us to perform an analysis of the effects on higher order correlators. 

In particular, using the invariance of the Chern-Simons term under a Weyl transformation of the metric, we showed that contributions to tensor-tensor-tensor $\Braket{ \gamma \gamma \gamma }$ and tensor-scalar-scalar $\Braket{ \gamma \zeta \zeta }$ bispectra are suppressed by the very small parameter $\Theta = (\pi/2) H/ M_{CS}$, where $H$ is the Hubble parameter during inflation and $M_{CS} = M^2_{Pl}/ (8 \dot{f})$ is the Chern-Simons mass, a characteristic energy scale at which instabilities may arise during inflation (here a dot stands for a derivative with respect to cosmic time). The quantity $\Theta$ must be very small during inflation ($\Theta \ll 1$) to avoid any instability (for more details on this parameter see ref.~\cite{Bartolo:2017szm}).

On the other hand, we showed explicitly that contributions to the bispectrum $\Braket{ \gamma \gamma \zeta }$ are not suppressed and we were able to define a parity breaking coefficient $\Pi$%
\footnote{The exact definition of $\Pi$ from ref.~\cite{Bartolo:2017szm} reads
  \begin{equation}
    \Pi = \frac{\braket{ \gamma_{R}(\vec{k})  \gamma_{R}(\vec{k})  \zeta(\vec{k}) }_{TOT} - \braket{ \gamma_{L}(-\vec{k})  \gamma_{L}(-\vec{k})  \zeta(-\vec{k}) }_{TOT}}{\braket{ \gamma_{R}(\vec{k})  \gamma_{R}(\vec{k})  \zeta(\vec{k}) }_{TOT} + \braket{ \gamma_{L}(-\vec{k})  \gamma_{L}(-\vec{k})  \zeta(-\vec{k}) }_{TOT}}
    = \frac{96\pi}{25} H^2 \frac{\partial^2 f(\phi)}{\partial \phi^2}\, .
  \end{equation}
}
of the bispectrum $\Braket{  \gamma \gamma \zeta }$ which turns out to be proportional to ${\partial^2 f(\phi)}/{\partial \phi^2}$, namely the second order derivative of the coupling $f(\phi)$ with respect to the scalar field (which is one of the reasons why the level of parity breaking in the bispectrum is not necessarily suppressed, unlike the power spectrum). 

In this paper we will provide a Fisher-matrix forecast about the detectability of this parameter $\Pi$ using $TTT$, $EEE$, $BBT$ and $BBE$ temperature (T) and polarization (E or B) cosmic microwave background (CMB) angular bispectra. This is the first time that CMB angular bispectra are used for making forecasts about parity violation in the $\Braket{ \gamma \gamma \zeta }$ bispectrum from an inflationary model (in this case from Chern-Simons gravity) and hence on the parameter $\Pi$.%
\footnote{
  For the detectability analysis on the parity-breaking $\Braket{ \gamma \gamma \zeta }$ bispectrum from primordial helical magnetic fields, see ref.~\cite{Shiraishi:2012sn}. On the other hand, there are studies on measuring parity violation in $\Braket{ \gamma \gamma}$ \cite{Gluscevic:2010vv}, $\Braket{ \gamma \gamma \gamma }$ \cite{Kamionkowski:2010rb,Shiraishi:2011st,Shiraishi:2012sn,Shiraishi:2013kxa,Namba:2015gja,Shiraishi:2016yun}, $\Braket{\gamma\zeta\zeta}$ \cite{Shiraishi:2012sn} and $\Braket{\zeta\zeta\zeta\zeta}$ \cite{Shiraishi:2016mok}. See refs.~\cite{Saito:2007kt,Gerbino:2016mqb} for the CMB bounds on chiral $\Braket{ \gamma \gamma}$ and refs.~\cite{Shiraishi:2014ila,Ade:2015ava} for those on chiral $\Braket{ \gamma \gamma \gamma}$. For an analysis on how to extract information on the chirality of gravitational waves from direct interferometric measurements see ref.~\cite{Bartolo:2016ami}.}
In particular, we will show that the minimum detectable value of $\Pi$ could be of order $10^6$ (see figure~\ref{fig:error_Pi} and eq.~\eqref{eq:dPi_BBT}). In ref.~\cite{Bartolo:2017szm} we showed that the assumption of an approximately constant $M_{CS}$ during inflation constrains the value of $\Pi$ to be not greater than the ratio $\Theta/\epsilon$, $\epsilon$ being the well-known slow roll parameter. In such a case it is not possible to detect $\Pi$, since in order to increase the value of $\Pi$  we should reduce the value of $\epsilon$ (or better to say $r$), causing at the same time the reduction of the S/N ratio on $\Pi$, as we can see in figure~\ref{fig:error_Pi} (for instance, if we take $r=10^{-2}$ and we assume $\Theta= 10^{-1}$, then we would get $\Pi<10^{2}$ and $\Delta \Pi = 10^{6}$, making the detection of $\Pi$ impossible). On the other hand, in the present paper we will show that one can easily generalize our results considering the time dependence of $M_{CS}$. In such a case any theoretical constraint over $\Pi$ vanishes. In particular, a model in which the Chern-Simons mass $M_{CS}$  varies significantly during inflation can drastically enhance $\Pi$ beyond the minimum detectable value. This result is promising since it offers a new way to test Chern-Simons inflation with a time dependent $M_{CS}$ in a regime where the power-spectrum statistics in any case can not provide any useful information. 

The paper is organized as follows. In section~\ref{sec:theory} we present the expression of the $\Braket{  \gamma \gamma \zeta }$  bispectrum arising from the Chern-Simons term and we will show how the parity-breaking parameter $\Pi$ can be enhanced by considering the time dependence of $M_{CS}$. In section~\ref{sec:CMB} we obtain the Fisher-matrix forecasts on the parameter $\Pi$ using $TTT$, $EEE$, $BBT$ and $BBE$ CMB angular bispectra. In section~\ref{sec:conclusions} we draw our conclusions.

\section{Enhancement of parity-odd non-Gaussianities in Chern-Simons gravity}\label{sec:theory}

To begin with, we very briefly recall how we get the Chern-Simons contribution to the tensor-tensor-scalar bispectrum $\Braket{  \gamma \gamma \zeta }$ (for more details on the computations see section~5 of ref.~\cite{Bartolo:2017szm}). The first step is to fix a gauge and evaluate the Chern-Simons term at third order in the primordial cosmological perturbations. The third-order interaction terms between 2 tensors and 1 scalar in the spatially flat gauge and at leading order in slow-roll read (in Fourier space) \cite{Bartolo:2017szm}
  \begin{eqnarray} \label{eq:L_stt}
  L_{int}^{\gamma \gamma \delta \phi}
  &\sim& \int d^3 k \int d^3 p \int d^3 q
  \frac{\delta^{(3)}(\vec k+ \vec p+ \vec q)}{(2 \pi)^6} \nonumber \\
  && \times \left\{ \lambda_s \left(\frac{\partial f(\phi)}{\partial \phi} \right) p \mbox{ } \delta \phi'(\vec k) \left[{\gamma'}^s_{ij}(\vec p) {\gamma'}^{s, \,ij}(\vec q) + \left( \vec{p} \cdot \vec{q} \right)\gamma^s_{ij} (\vec p) \gamma^{s, \,ij} (\vec q)\right] \right. \nonumber \\
&& \left. \quad + \lambda_s \, a \left(\dot \phi \frac{\partial^2 f(\phi)}{\partial^2 \phi} \right) p \mbox{ }\delta \phi(\vec k) \left[{\gamma'}^s_{ij}(\vec p) {\gamma'}^{s, \,ij}(\vec q) +  \left( \vec{p} \cdot \vec{q} \right)\gamma^s_{ij} (\vec p) \gamma^{s,\, ij} (\vec q)  \right] \right\} .
\end{eqnarray}
Here ${\gamma}^s_{ij}(\vec p)={\gamma}_s(\vec p) \, {\epsilon}^s_{ij}(\vec p)$, where ${\epsilon}^s_{ij}(\vec p)$ is the polarization tensor and ${\gamma}_s(\vec p)$ is the tensor mode function. The latin indices contractions are made with $\delta^{ij}$ and the primes $'$ indicate derivatives with respect to conformal time. The coefficient $\lambda_s$ takes $+1$ ($-1$) for R (L) polarization modes and the sum over the polarization index $s=R,L$ is understood for simplicity of notation.

In the second line of the Lagrangian \eqref{eq:L_stt} there are interaction vertices that depend on the second  derivative of the coupling $f(\phi)$ with respect to the inflaton field, while in the first line the interaction terms depend on the first derivative of the coupling. The latter turns out to be proportional to the ratio $H/M_{CS}$, thus they are highly  suppressed and can be ignored. The remaining terms give the relevant contribution to the tensor-tensor-scalar bispectrum $\Braket{  \gamma \gamma \delta \phi }$. The computation of the bispectrum is made through the so-called In-In formalism (see e.g. refs.~\cite{Weinberg:2005,Maldacena:2003, Collins:2011, Chen:2007})
\begin{equation}
  \Braket{ \gamma_{s}(\vec{k}_1) \gamma_{s'}({\vec k}_2) \delta \phi({\vec k}_3) }
  = - i \int_{-\infty}^{0} d\tilde{\tau}
  \Braket{ 0 | \left[\gamma_{s}(\vec{k}_1, 0)\gamma_{s'}(\vec{k}_2, 0) \delta \phi(\vec{k}_3, 0), \, H^{\gamma \gamma \delta \phi}_{int}(\tilde{\tau}) \right]  | 0 } , \label{master}
\end{equation}
where $H^{\gamma \gamma \delta \phi}_{int} = - L^{\gamma \gamma \delta \phi}_{int}$ is the cubic interaction Hamiltonian. The final step to obtain $\Braket{  \gamma \gamma \zeta }$ is to pass from the inflaton perturbation variable $\delta \phi$ to the gauge invariant curvature perturbation $\zeta$ through the following non-linear relation on super-horizon scales \cite{Maldacena:2003} 
\begin{equation}
\zeta  = \zeta_1 + \frac{1}{2} \frac{\ddot{\phi}}{\dot{\phi} H} \zeta_1^2 + \frac{1}{4} \frac{\dot{\phi}^2}{H^2} \zeta_1^2 \, , \label{zeta_non_linear}
\end{equation} 
where $\phi$ is the background value of the inflaton field and 
\begin{equation}
\zeta_1= - \frac{H}{\dot \phi} \delta \phi \, .
\end{equation}
The final bottom-line expression of the tensor-tensor-scalar bispectrum $\Braket{  \gamma \gamma \zeta }$ reads~\cite{Bartolo:2017szm} (see, also, for an equivalent expression eq.~(5.62) of ref.~\cite{Bartolo:2017szm})
\begin{eqnarray}
\begin{split}
\Braket{\gamma_{R/L}(\vec{k}_1) \gamma_{R/L}(\vec{k}_2) \zeta(\vec{k}_3) }
&= (2\pi)^3 \delta^{(3)}\left(\sum_{n = 1}^3 \vec{k}_n\right) \\
&\quad \times
(+/-) F_{k_1 k_2 k_3} \,  (\hat{k}_1 \cdot \hat{k}_2) \left[ \epsilon_{ij}^{R/L}(\vec{k}_1) \epsilon_{ij}^{R/L}(\vec{k}_2) \right]^* \,, \\  
\Braket{\gamma_{R/L}(\vec{k}_1) \gamma_{L/R}(\vec{k}_2) \zeta(\vec{k}_3)} &= 0 \,, \label{eq:hhzeta}
\end{split}
\end{eqnarray}
where $\epsilon_{ij}^{R/L}(\vec{k})$ is the polarization tensor of R(L)-handed gravitational waves,
\begin{eqnarray} 
F_{k_1 k_2 k_3} &=& - \frac{\pi H^6}{2 M_{Pl}^4} \frac{\partial^2 f(\phi)}{\partial \phi^2} 
\frac{(k_1 + k_2)}{k_1^2 k_2^2 k_3^3}
= - 
\frac{25 \pi^4}{768} {\cal P}_\zeta^2 \left( r^2 \Pi \right)
\frac{(k_1 + k_2)}{k_1^2 k_2^2 k_3^3} \, , \label{eq:Fkkk} 
\end{eqnarray}
where we have used 
\begin{eqnarray} 
\left(\frac{H}{M_{Pl}}\right)^2 &=& \frac{\pi^2}{2} r {\cal P}_\zeta \, , \label{eq:power}
\end{eqnarray}
and 
\begin{equation} \label{eq:pi_value}
\Pi =\frac{96\pi}{25} H^2 \frac{\partial^2 f(\phi)}{\partial \phi^2} \, ,
\end{equation}
$r$ being the tensor-to-scalar ratio. Equation~\eqref{eq:power} holds since the standard relation $r = 16 \epsilon$ is still valid in the current model apart for a very small correction in the total tensor power spectrum which is negligible in the $\Theta \ll 1$ regime (see eq.~(4.23) of ref.~\cite{Bartolo:2017szm}). According to the results of ref.~\cite{Bartolo:2017szm}, the bispectrum \eqref{eq:hhzeta} peaks in the squeezed configurations $k_1 \sim k_2 \gg k_3$. As we will show  in the next section, this fact will lead to a signal-to-noise ratio enhancement when using $BBT$ or $BBE$ CMB bispectra to make a measurement of the parameter $\Pi$. 

Moreover, in ref.~\cite{Bartolo:2017szm} we derived a theoretical constraint such that the parameter $\Pi$ must be smaller than the ratio $\Theta/\epsilon$. This constraints follows by considering the Chern-Simons mass as approximately constant in time during inflation. This assumption is made for solving analytically the equations of motion of tensor mode functions in terms of Whittaker functions and providing a general solution for the super-horizon tensor power spectra. However, let us briefly discuss in the following how we can definitely drop such a constraint, treating $\Pi$ essentially as a free parameter. 

For this purpose, let us consider the quadratic Lagrangian for tensor modes, i.e. eq.~(4.5) of ref.~\cite{Bartolo:2017szm}
\begin{equation} \label{eq:action_tt2}
S|_{\gamma \gamma}= \sum_{s=L, R} \int d\tau \mbox{ }\frac{d^3 k}{(2 \pi)^3} \mbox{  } A^2_{T, s} \left[\mbox{ }{|\gamma'_s (\tau,k)|}^2-k^2 {|\gamma_s (\tau,k)|}^2 \mbox{ }\right] , 
\end{equation}
where
\begin{equation}\label{eq:A}
A^2_{T, s}= \frac{M^2_{Pl}}{2} a^2 \left(1 - 8 \lambda_s  \frac{k}{a} \frac{\dot f(\phi)}{M^2_{Pl}} \right) = \frac{M^2_{Pl}}{2} a^2 \left(1 - \lambda_s  \frac{k_{phys}}{M_{CS}}  \right)\, ,
\end{equation}
with  
\begin{equation}\label{eq:mass_CS}
M_{CS}= \frac{ M^2_{Pl}}{ 8 \dot f(\phi)} \, .
\end{equation}	
Following ref.~\cite{Bartolo:2017szm}, let us employ the field redefinition
\begin{equation}\label{eq:redefinition}
\mu_s= A_{T, s} \gamma_s \, ,
\end{equation}		
so that the action becomes 
\begin{equation} 
S|_{\gamma \gamma}= \sum_{s=L, R} \int d\tau \mbox{ }\frac{d^3 k}{(2 \pi)^3} \mbox{  }  \left[\mbox{ }{|\mu'_s (\tau,k)|}^2-k^2 {|\mu_s (\tau,k)|^2}+ \frac{A''_{T, s}}{A_{T, s}}|\mu_s (\tau,k)|^2 \mbox{ }\right] \, .
\end{equation}	
The equations of motion for $\mu_s$ read
\begin{equation} \label{eq:eom}
\mu''_s (\tau,k)+ \left(k^2 - \frac{A''_{T, s}}{A_{T, s}} \right) \mu_s (\tau,k) = 0 \, ,
\end{equation}
where, apart from slow-roll corrections, the exact effective mass term reads
\begin{equation} \label{eq:mass_term}
\frac{A''_{T, s}}{A_{T, s}} = \frac{2}{\tau^2} \left(1- \frac{\lambda_s}{2} k \tau \frac{H}{M_{CS}} \mathcal{A}\right)=\frac{2}{\tau^2} \left(1+\frac{\lambda_s}{2} \frac{k_{phys}}{H} \frac{H}{M_{CS}} \mathcal{A}\right)\, .
\end{equation}
In eq.~\eqref{eq:mass_term} we have defined the following quantity
\begin{equation}
\mathcal{A} = \frac{1}{(1-\lambda_s k_{phys}/M_{CS})^2} \left\{ \left[1- \xi + \frac{\omega}{2}  - \frac{\xi}{2H \tau} 
\right] \left(1-\lambda_s \frac{k_{pyhs}}{M_{CS}}\right) - \lambda_s \frac{k_{phys}}{2 M_{CS}} \left[ \frac{1}{2} + \xi + \frac{\xi^2 }{2} \right]\right\}
\end{equation}
where \cite{Bartolo:2017szm}
\begin{eqnarray}
k_{phys} &=& \frac{k}{a} \, , \\
\xi &=& \frac{\dot M_{CS}}{M_{CS} H} \, , \label{eq:xi}
\end{eqnarray}
and
\begin{equation}
\omega = \frac{\ddot M_{CS}}{M_{CS} H^2} \, ,
\end{equation}
where the dots stand for derivatives with respect to cosmological time $t$.

We can notice that if we neglect the time dependence of $M_{CS}$ (i.e. $\xi, \omega \ll 1$) and we consider that during inflation we need $k_{phys}/M_{CS} \ll 1$ and $H/M_{CS} \ll 1$ in order to avoid instabilities (see ref.~\cite{Bartolo:2017szm}), then $\mathcal{A} \simeq 1$. In this case we perfectly recover eq.~(4.9) of ref.~\cite{Bartolo:2017szm} (apart from slow-roll corrections). 

However, let us assume that $M_{CS}$ is not nearly constant and it is fast growing%
\footnote{A-priori $M_{CS}$ can also decrease in time, but in this case we could face some instabilities since at a certain time the ratio $H/M_{CS}$ might become order unity. For this reason it is safer to consider a scenario in which $M_{CS}$ increases.}
in time during inflation. In this case  the $\xi$, $\omega$ parameters can also be much greater than 1 during inflation and time dependent. Looking at eq.~\eqref{eq:mass_term} we have that at the beginning of inflation, when a comoving mode $k$ is well-inside the Hubble horizon (this corresponds to the limit $- k \tau = + \infty$), the effective mass term is negligible and it starts to be important just at later times (when $-k \tau \rightarrow 0$). But, if initially $k_{phys}/M_{CS} \ll ~1$, $H/M_{CS} \ll 1$ and $M_{CS}$ grows fast enough during inflation, then regardless how much $\xi$ and $\omega$ are large the factor $H/M_{CS}$ washes away soon any parity breaking term in the effective mass \eqref{eq:mass_term} and hence in the tensor power spectra.%
\footnote{We can understand this fact also directly from eq.~\eqref{eq:A}. If at the beginning of inflation we assume $k_{phys}/M_{CS} \ll ~1$ and during inflation $M_{CS}$ increases in time, then the coefficient  $A_{T, s}^2$ becomes equal to $M^2_{Pl} a^2/2$ plus a very small correction which can be neglected in studying the dynamics of $\gamma_s$.}
In such a scenario, the net effect is that any parity-breaking signature in the tensor primordial power spectrum on super-horizon scales is expected to be extremely small. Notice that now one should solve the equation of motion \eqref{eq:eom} in a way completely different from what described in ref.~\cite{Bartolo:2017szm}, since it is no more a Whittaker equation. Anyway, here we are not interested in finding a precise solution for the tensor power spectrum, since, as we just explained, the corrections to the standard gravity case are expected to be negligible. 
	
On the contrary, the fact that the Chern-Simons mass is time dependent, now does release the parity-breaking level $\Pi$ of the bispectrum from the theoretical constraints found in ref.~\cite{Bartolo:2017szm}, since higher-order correlators whose amplitudes are sensitive only to the time dependence of the Chern-Simons mass may become unconstrained. Let us explain this fact quantitatively. Using the definition of $M_{CS}$, eq.~\eqref{eq:mass_CS}, it is possible to rewrite the parameter $\xi$ in eq.~\eqref{eq:xi} as%
\footnote{In ref.~\cite{Bartolo:2017szm} the parameter $\xi$ in eq.~(5.73) has the + sign because of a typo.}
\begin{equation}\label{eq:value_xi}
  \xi = - \sqrt{2 \epsilon} M_{Pl}  \frac{\partial^2 f(\phi)/\partial \phi^2}{\partial f(\phi)/\partial \phi} \, ,
\end{equation}
where we have neglected some slow-roll corrections. Equation~\eqref{eq:value_xi} can be rewritten in terms of $\Pi$ using the definition \eqref{eq:pi_value} 
\begin{equation} \label{eq:xi2}
  \xi = - \frac{25}{96 \pi}\sqrt{2 \epsilon} \frac{M_{Pl}}{H}  \frac{\Pi}{H \, \partial f(\phi)/\partial \phi} \, .
\end{equation}
From this last equation we see that the time dependence of the Chern-Simons mass $M_{CS}$ is directly related to the parameter $\Pi$, i.e. to the higher order derivatives of the coupling function $f(\phi)$. In particular, using eq. \eqref{eq:mass_CS} and recalling the definition of $\Theta$
\begin{equation}
\Theta = \frac{\pi}{2} \frac{H}{M_{CS}} \, ,
\end{equation}	
we can rewrite eq.~\eqref{eq:xi2} as
\begin{equation} 
\Pi = - \frac{12}{25} \frac{\Theta}{\epsilon} \xi \, .
\end{equation}
Now, since the parameter $\xi$ controls the time dependence of the Chern-Simons mass, if we assume that this mass is almost constant in time during inflation, then $|\xi|$ must be not larger than 1. As a consequence, we get the theoretical constraint
\begin{equation} 
|\Pi|  \lesssim \frac{12}{25} \frac{\Theta}{\epsilon} \, .
\end{equation}
 On the contrary, if we do not have any strong constraint on the time variation of $M_{CS}$, we can treat $\xi$ (and hence $\Pi$) as a free parameter and use the observations to constrain how much the Chern-Simons mass could have increased in time during inflation.
Since $\Pi$ is a coefficient appearing in the explicit expression of  $\Braket{ \gamma \gamma \zeta }$  bispectrum, eq.~\eqref{eq:hhzeta}, this makes the latter observable a good candidate for testing Chern-Simons gravity admitting a Chern-Simons mass which is growing during inflationary epoch.

\section{Signatures in the CMB bispectra}\label{sec:CMB}

In this section, we explore the signatures of the parity-odd tensor-tensor-scalar bispectrum discussed in the previous section in the CMB temperature and polarization bispectra. 

\subsection{Notations}

For CMB bispectrum computations, we follow the procedure developed in refs.~\cite{Shiraishi:2010sm,Shiraishi:2010kd}. We therefore have to express the tensor-tensor-scalar bispectrum with the polarization tensors used in ref.~\cite{Shiraishi:2010kd}, satisfying $e_{ij}^{(+2 / -2)}(\hat{k}) = - \epsilon_{ij}^{R/L}(\vec{k})$. Since we now consider the tensor-mode decomposition using
\begin{eqnarray}
 \gamma_{ij}(\vec{x}) = \int \frac{d^3 \vec{k}}{(2\pi)^3}
 e^{i \vec{k} \cdot \vec{x}} \sum_{\lambda = \pm 2} \gamma_{\vec{k}}^{(\lambda)} e_{ij}^{(\lambda)}(\hat{k}) ~,
\end{eqnarray}
we have $\gamma_{\vec{k}}^{(+2 / -2)} = - \gamma_{R/L}(\vec{k})$. Also, changing the notation of $\zeta (\vec{k}_3)$ to $\zeta_{\vec{k}_3}$ for simplicity, eq.~\eqref{eq:hhzeta} can be rewritten as
\begin{equation}
\Braket{\gamma_{\vec{k}_1}^{(\lambda_1)} \gamma_{\vec{k}_2}^{(\lambda_2)} \zeta_{\vec{k}_3}}
  = (2\pi)^3 \delta^{(3)}\left(\sum_{n = 1}^3 \vec{k}_n\right)
  F_{k_1 k_2 k_3}  \left(\frac{\lambda_1}{2} \right) \delta_{\lambda_1, \lambda_2} (\hat{k}_1 \cdot \hat{k}_2) e_{ij}^{(-\lambda_1)}(\hat{k}_1) e_{ij}^{(-\lambda_2)}(\hat{k}_2) \, . \label{eq:hhzeta_CMB}
\end{equation}
The spherical harmonic coefficients of the temperature ($X = T$) and E/B-mode polarization ($X = E/B$) anisotropies from the scalar ($\zeta$) and the tensor ($\gamma^{(\pm 2)}$), are expressed, respectively, as \cite{Shiraishi:2010sm,Shiraishi:2010kd}
\begin{eqnarray}
  a_{\ell m}^{(s) X} &=& 
4\pi (-i)^{\ell} \int \frac{d^3 \vec{k}}{(2\pi)^{3}}
{\cal T}_{\ell(s)}^{X}(k) \zeta_{\vec{k}}  Y_{\ell m}^*(\hat{k}) ~, \\
  a_{\ell m}^{(t) X} &=& 
4\pi (-i)^{\ell} \int \frac{d^3 \vec{k}}{(2\pi)^{3}}
{\cal T}_{\ell(t)}^{X}(k)  \sum_{\lambda = \pm 2} \left(\frac{\lambda}{2}\right)^x \gamma_{\vec{k}}^{(\lambda)} {}_{-\lambda} Y_{\ell m}^*(\hat{k}) ~, 
\end{eqnarray}
where ${}_{\lambda} Y_{\ell m}(\hat{k})$ is the spin-weighted spherical harmonic function, ${\cal T}_{\ell (s)}^{X}(k)$ and ${\cal T}_{\ell (t)}^{X}(k)$ is the scalar and tensor transfer functions, respectively, and $x$ takes $0$ ($1$) for $X = T, E$ ($X = B$). Using these, the CMB bispectra sourced by the primordial tensor-tensor-scalar correlators can be written as 
\begin{eqnarray}
  \Braket{a_{\ell_1 m_1}^{(t) X_1} a_{\ell_2 m_2}^{(t) X_2} a_{\ell_3 m_3}^{(s) X_3} }
  &=& \left[ \prod_{n=1}^2 4\pi (-i)^{\ell_n} \int \frac{d^3 \vec{k}_n}{(2\pi)^{3}}
{\cal T}_{\ell_n (t)}^{X_n}(k_n)  \sum_{\lambda_n = \pm 2} \left(\frac{\lambda_n}{2}\right)^{x_n}  {}_{-\lambda_n} Y_{\ell_n m_n}^*(\hat{k}_n) \right]
\nonumber \\ 
&&  \times \, 4\pi (-i)^{\ell_3} \int \frac{d^3 \vec{k}_3}{(2\pi)^{3}}
 {\cal T}_{\ell_3 (s)}^{X_3}(k_3)   Y_{\ell_3 m_3}^*(\hat{k}_3)
 \Braket{\gamma_{\vec{k}_1}^{(\lambda_1)} \gamma_{\vec{k}_2}^{(\lambda_2)} \zeta_{\vec{k}_3}} \, . \label{eq:CMB_bis_ini}
 \end{eqnarray}

\subsection{Allowed harmonic-space configurations}

We, at first, check the $\ell$-space configurations where nonvanishing signal lies in by following the procedure discussed in ref.~\cite{Shiraishi:2016ads}.

Using the fact that ${}_{-\lambda} Y_{\ell m}(-\hat{k}) = (-1)^\ell {}_{\lambda} Y_{\ell m}(\hat{k})$, we can rewrite eq.~\eqref{eq:CMB_bis_ini} as 
\begin{eqnarray}
  \Braket{a_{\ell_1 m_1}^{(t) X_1} a_{\ell_2 m_2}^{(t) X_2} a_{\ell_3 m_3}^{(s) X_3} }
  &=& \left[ \prod_{n=1}^2 4\pi (-i)^{\ell_n} \int \frac{d^3 \vec{k}_n}{(2\pi)^{3}}
{\cal T}_{\ell_n (t)}^{X_n}(k_n)  \sum_{\lambda_n = \pm 2} \left(\frac{\lambda_n}{2}\right)^{x_n}  {}_{-\lambda_n} Y_{\ell_n m_n}^*(\hat{k}_n) \right]
\nonumber \\ 
&& \times \, 4\pi (-i)^{\ell_3} \int \frac{d^3 \vec{k}_3}{(2\pi)^{3}}
{\cal T}_{\ell_3 (s)}^{X_3}(k_3) Y_{\ell_3 m_3}^*(\hat{k}_3) \nonumber \\ 
&& \times \, (-1)^{x_1 + x_2 + \ell_1+ \ell_2 + \ell_3 } \Braket{\gamma_{- \vec{k}_1}^{(-\lambda_1)} \gamma_{- \vec{k}_2}^{(-\lambda_2)} \zeta_{- \vec{k}_3}}\, . \label{eq:CMB_bis_ini2}
 \end{eqnarray}
The tensor-tensor-scalar bispectrum under consideration \eqref{eq:hhzeta_CMB} has odd parity; namely, it obeys
\begin{eqnarray}
  \Braket{\gamma_{\vec{k}_1}^{(\lambda_1)} \gamma_{\vec{k}_2}^{(\lambda_2)} \zeta_{\vec{k}_3}} = - \Braket{\gamma_{- \vec{k}_1}^{(-\lambda_1)} \gamma_{- \vec{k}_2}^{(-\lambda_2)} \zeta_{- \vec{k}_3}}. \label{eq:hhzeta_parity} 
\end{eqnarray}
Comparing eq.~\eqref{eq:CMB_bis_ini} with eq.~\eqref{eq:CMB_bis_ini2} under the parity-odd condition \eqref{eq:hhzeta_parity}, we find that
\begin{eqnarray}
  \Braket{a_{\ell_1 m_1}^{(t) X_1} a_{\ell_2 m_2}^{(t) X_2} a_{\ell_3 m_3}^{(s) X_3} }[1 + (-1)^{x_1 + x_2 + \ell_1+ \ell_2 + \ell_3 }] = 0
\end{eqnarray}
always holds; thus, nonvanishing signal is confined to%
\footnote{Notice that this statement is perfectly in agreement with and generalizes what is qualitatively already mentioned in the conclusions of ref.~\cite{Bartolo:2017szm}.}
\begin{eqnarray}
x_1 + x_2 + \ell_1+ \ell_2 + \ell_3 = {\rm odd}. \label{eq:P-odd_rule}
\end{eqnarray}
This means that nonvanishing signal in $X_1 X_2 X_3$ and $BBX_3$ ($B X_2 X_3$ and $X_1 B X_3$), where $X_1, X_2, X_3 = T, E$, arises from odd (even) $\ell_1 + \ell_2 + \ell_3$ components. It is worth stressing that these combinations are not realized under the usual parity-conserving theories like Einstein gravity and therefore they can become distinctive indicators of Chern-Simons gravity if they are detected. The restriction given by eq.~\eqref{eq:P-odd_rule} is, of course, confirmed also in the following CMB bispectrum formulation.

\subsection{CMB bispectrum formulation}

Plugging eq.~\eqref{eq:hhzeta_CMB} into eq.~\eqref{eq:CMB_bis_ini} yields
\begin{eqnarray}
 \Braket{a_{\ell_1 m_1}^{(t) X_1} a_{\ell_2 m_2}^{(t) X_2} a_{\ell_3 m_3}^{(s) X_3} }
  &=& \left[ \prod_{n=1}^3 \frac{(-i)^{\ell_n}}{\pi} \int_0^\infty k_n^2 d k_n 
  \int d^2 \hat{k}_n \right] {\cal T}_{\ell_1 (t)}^{X_1}(k_1) {\cal T}_{\ell_2 (t)}^{X_2}(k_2)  {\cal T}_{\ell_3 (s)}^{X_3}(k_3)
 \nonumber \\
 && \times \, \sum_{\lambda_1 = \pm 2} \left(\frac{\lambda_1}{2}\right)^{x_1 + x_2 + 1}
 {}_{-\lambda_1} Y_{\ell_1 m_1}^*(\hat{k}_1)  {}_{-\lambda_1} Y_{\ell_2 m_2}^*(\hat{k}_2) Y_{\ell_3 m_3}^*(\hat{k}_3)
 \nonumber \\
&& \times \, \delta^{(3)}\left(\sum_{n = 1}^3 \vec{k}_n\right) F_{k_1 k_2 k_3} \,
    (\hat{k}_1 \cdot \hat{k}_2) \, e_{ij}^{(-\lambda_1)}(\hat{k}_1) e_{ij}^{(-\lambda_1)}(\hat{k}_2)\, .
\end{eqnarray}
Here, the angular-dependent parts are decomposed using the spin-weighted spherical harmonics as
\begin{eqnarray}
 (\hat{k}_1 \cdot \hat{k}_2) e_{ij}^{(-\lambda_1)}(\hat{k}_1) e_{ij}^{(-\lambda_1)}(\hat{k}_2)
     &=& \frac{2^5 \pi^2}{15}     
  \sum_{j \mu} {}_{\lambda_1}Y_{j \mu}^*(\hat{k}_1)  
      {}_{\lambda_1}Y_{j -\mu}^*(\hat{k}_2)  (-1)^{\mu + 1 + j}
      \frac{\left( h_{1 ~ 2 ~j}^{0 \lambda_1 -\lambda_1} \right)^2 }{2j + 1} , \\
      \delta^{(3)}\left( \sum_{n=1}^3 \vec{k}_n  \right) 
      &=& 8 \int_0^\infty y^2 dy
      \left[ \prod_{n=1}^3 \sum_{L_n M_n} 
 (-1)^{\frac{L_n}{2}} j_{L_n}(k_n y) 
Y_{L_n M_n}^*(\hat{k}_n) \right] \nonumber \\
&& \times \, \left(
  \begin{array}{ccc}
  L_1 & L_2 & L_3 \\
  M_1 & M_2 & M_3 
  \end{array}
  \right)
  h_{L_1 L_2 L_3}^{0 ~ 0 ~ 0} \, ,
 \end{eqnarray}
where
\begin{eqnarray}
h^{s_1 s_2 s_3}_{l_1 l_2 l_3}
\equiv \sqrt{\frac{(2 l_1 + 1)(2 l_2 + 1)(2 l_3 + 1)}{4 \pi}}
\left(
  \begin{array}{ccc}
  l_1 & l_2 & l_3 \\
  s_1 & s_2 & s_3
  \end{array}
  \right)
\end{eqnarray}
with
$\left(
\begin{array}{ccc}
  a & b & c \\
  d & e & f
\end{array}
\right)$ the Wigner $3j$ symbol. The products of the resulting (spin-weighted) spherical harmonics are integrated with respect to $\hat{k}_n$ according to the identities: 
  \begin{eqnarray}
     \int d^2 \hat{k}  \prod_{n = 1}^2 Y_{l_n m_n}^*(\hat{k}) &=& (-1)^{m_1}\delta_{l_1, l_2} \delta_{m_1, -m_2}\,  , \\
     \int d^2 \hat{k} \prod_{n = 1}^3 {}_{s_n} Y_{l_n m_n}^*(\hat{k}) &=& h_{~l_1 ~~ l_2 ~~ l_3}^{-s_1 -s_2 -s_3}
  \left(
  \begin{array}{ccc}
  l_1 & l_2 & l_3 \\
  m_1 & m_2 & m_3 
  \end{array}
  \right) \, .
  \end{eqnarray}
  Adding the angular momenta in the induced Wigner symbols by use of, e.g., 
\begin{eqnarray}
&& \sum_{m_4 m_5 m_6} (-1)^{\sum_{n=4}^6( l_n - m_n) }
\left(
\begin{array}{ccc}
  l_5 & l_1 & l_6 \\
  m_5 & -m_1 & -m_6 
 \end{array}
 \right) 
\left(
\begin{array}{ccc}
  l_6 & l_2 & l_4 \\
  m_6 & -m_2 & -m_4 
\end{array}
  \right)
\left(
 \begin{array}{ccc}
  l_4 & l_3 & l_5 \\
  m_4 & -m_3 & -m_5 
\end{array}
 \right) \nonumber \\
 &&\qquad\qquad
 = \left(
  \begin{array}{ccc}
  l_1 & l_2 & l_3 \\
  m_1 & m_2 & m_3 
 \end{array}
 \right) 
\left\{
 \begin{array}{ccc}
  l_1 & l_2 & l_3 \\
  l_4 & l_5 & l_6 
 \end{array}
 \right\} \, , 
\end{eqnarray}
where 
  $\left\{
  \begin{array}{ccc}
  a & b & c \\
  d & e & f
  \end{array}
  \right\}$
  is the Wigner $6j$ symbol, and performing the summations with respect to $\lambda_1$, we obtain the bottom-line expression $\Braket{a_{\ell_1 m_1}^{(t) X_1} a_{\ell_2 m_2}^{(t) X_2} a_{\ell_3 m_3}^{(s) X_3} } = B_{(tts) \ell_1 \ell_2 \ell_3}^{X_1 X_2 X_3}
 \left(
  \begin{array}{ccc}
  \ell_1 & \ell_2 & \ell_3 \\
  m_1 & m_2 & m_3 
  \end{array}
  \right)  $ where the angle-averaged bispectrum turns out to be 
\begin{eqnarray} \label{eq:forecast_tts}
  B_{(tts) \ell_1 \ell_2 \ell_3}^{X_1 X_2 X_3} &=&   \delta_{x_1+ x_2 + \ell_1 + \ell_2 + \ell_3}^{\rm odd}  
  (-i)^{\ell_1 + \ell_2 + \ell_3} \sum_{L_1 L_2}  (-1)^{\frac{L_1 + L_2 + \ell_3}{2}} h_{L_1 L_2 \ell_3}  \nonumber \\
 &&  \times \, \frac{2^6 \pi ^2}{15}     
 \sum_{j} 
 \frac{(-1)^{1 + j + L_2 + \ell_1}}{2j + 1}
 h_{L_1 \ell_1 j}^{0 2 -2} h_{L_2 \ell_2 j}^{0 2 -2} \left( h_{1 ~2 ~ j}^{0 2 -2} \right)^2
    \left\{
  \begin{array}{ccc}
  \ell_1 & \ell_2 & \ell_3 \\
  L_2 & L_1 & j 
  \end{array}
  \right\} \nonumber \\
 &&  \times \, \int_0^\infty y^2 dy 
 \left[ \prod_{n=1}^2 \frac{2}{\pi}  \int_0^\infty k_n^2 dk_n
   {\cal T}_{\ell_n (t)}^{X_n}(k_n) j_{L_n}(k_n y) \right] \nonumber \\
&&  \times \, \frac{2}{\pi} \int_0^\infty k_3^2 d k_3 {\cal T}_{\ell_3 (s)}^{X_3}(k_3)  j_{\ell_3}(k_3 y) 
 F_{k_1 k_2 k_3} \, ,
\end{eqnarray}
with
\begin{equation}
  \delta_{\ell}^{\rm odd} = \begin{cases} 1 & (\ell = {\rm odd}) \\
    0 & (\ell = {\rm even}) \end{cases} \, .
\end{equation}
In fact the ranges of the summations with respect to $L_1$, $L_2$ and $j$ are limited to a few modes by the selection rules of the Wigner symbols.

\subsection{Fisher matrix forecasts}

\begin{figure}[t!]
  \begin{tabular}{cc}
\begin{minipage}{0.5\hsize}
  \begin{center}
    \includegraphics[width=1\textwidth]{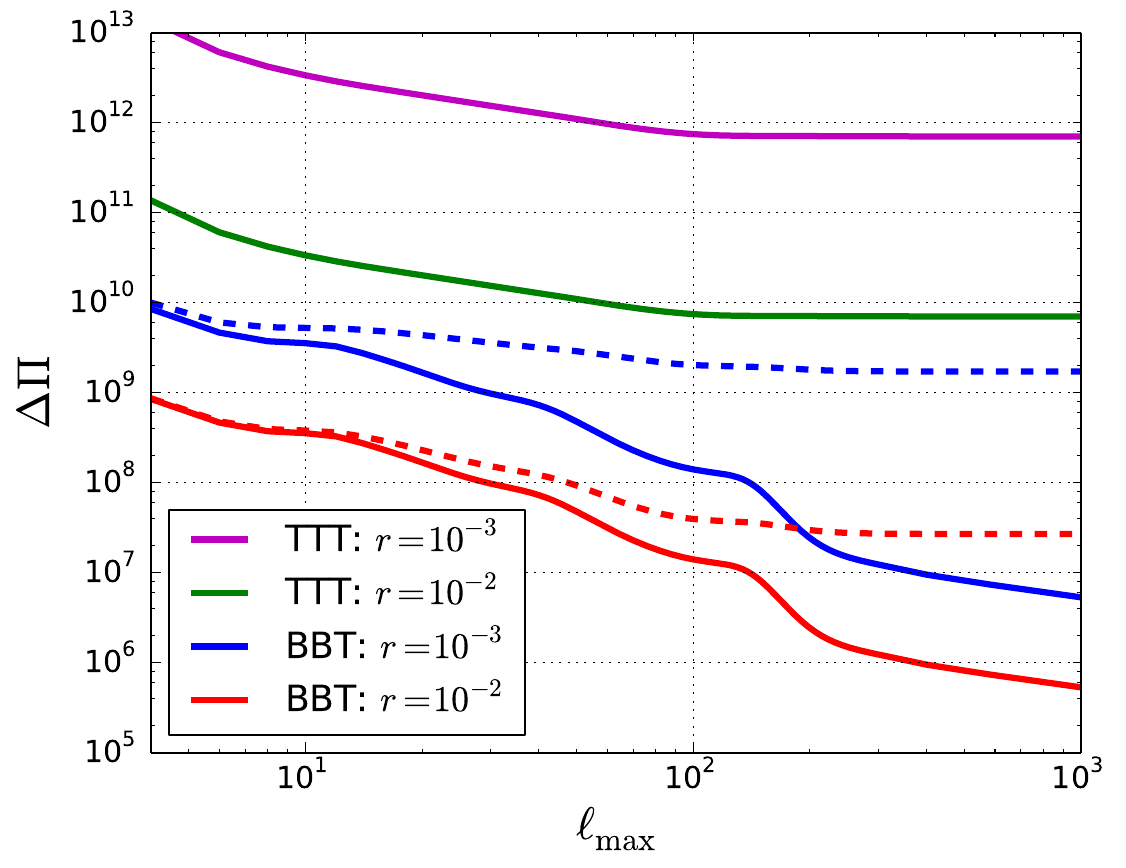}
  \end{center}
\end{minipage}
\begin{minipage}{0.5\hsize}
  \begin{center}
    \includegraphics[width=1\textwidth]{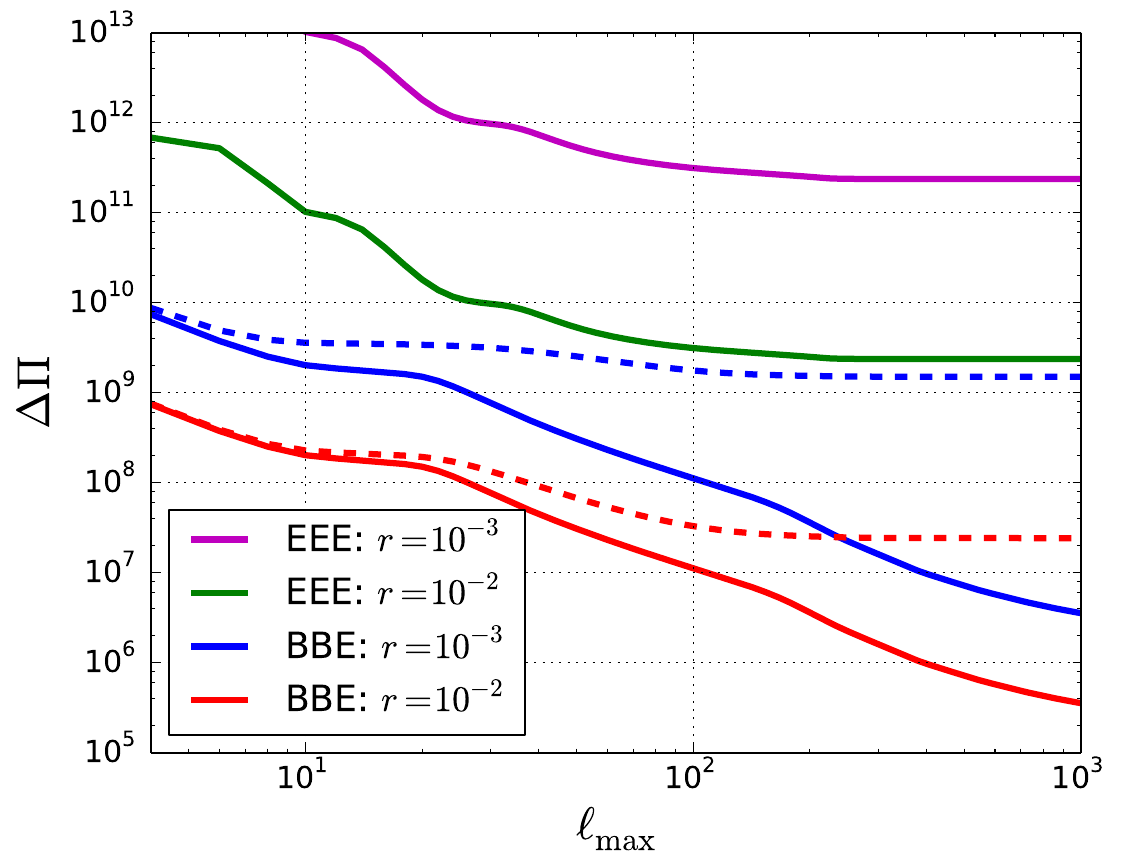}
  \end{center}
\end{minipage}
  \end{tabular}
\caption{Expected $1\sigma$ errors on the nonlinear chirality parameter $\Pi$ from $TTT$, $EEE$, $BBT$ and $BBE$ for $r = 10^{-2}$ and $10^{-3}$ as a function of $\ell_{\rm max}$. Here we assume a full-sky measurement without any instrumental uncertainties. Solid (dashed) lines for $BBT$ and $BBE$ are computed assuming a perfectly delensed (undelensed) B-mode polarization data.} \label{fig:error_Pi}
\end{figure}


From eq.~\eqref{eq:Fkkk}, one can see that $F_{k_1 k_2 k_3}$ depends linearly on the chirality parameter $\Pi$, given by eq.~\eqref{eq:pi_value}, and quadratically on the tensor-to-scalar ratio $r$; thus, $B_{(tts) \ell_1 \ell_2 \ell_3}^{X_1 X_2 X_3} \propto r^2 \Pi$.

  We now evaluate the detectability of $\Pi$ by computing the Fisher matrix from $XXX$ (${\cal F}^{XXX}$) or $BBX$ (${\cal F}^{BBX}$), with $X = T, E$, according to
\begin{eqnarray}
  {\cal F}^{XXX} &=& \sum_{\ell_1, \ell_2, \ell_3 = 2}^{\ell_{\rm max}}
   \frac{\left( \hat{B}_{\ell_1 \ell_2 \ell_3}^{XXX} \right)^2 }{6 C_{\ell_1}^{XX} C_{\ell_2}^{XX} C_{\ell_3}^{XX}} (-1)^{\ell_1 + \ell_2 + \ell_3} \, , \\
   {\cal F}^{BBX} &=& \sum_{\ell_1, \ell_2, \ell_3 = 2}^{\ell_{\rm max}}
   \frac{\left( \hat{B}_{\ell_1 \ell_2 \ell_3}^{BBX} \right)^2 }{2 C_{\ell_1}^{BB} C_{\ell_2}^{BB} C_{\ell_3}^{XX}} (-1)^{\ell_1 + \ell_2 + \ell_3} \, , \label{eq:fish_BBX}
 \end{eqnarray}
where $\hat{B}_{\ell_1 \ell_2 \ell_3} \equiv \partial B_{\ell_1 \ell_2 \ell_3} / \partial \Pi =  B_{\ell_1 \ell_2 \ell_3} / \Pi$, $B_{\ell_1 \ell_2 \ell_3}^{XXX} \equiv B_{(tts) \ell_1 \ell_2 \ell_3}^{XXX} + B_{(tst) \ell_1 \ell_2 \ell_3}^{XXX} + B_{(stt) \ell_1 \ell_2 \ell_3}^{XXX} = B_{(tts) \ell_1 \ell_2 \ell_3}^{XXX} + B_{(tts) \ell_3 \ell_1 \ell_2}^{XXX} + B_{(tts) \ell_2 \ell_3 \ell_1}^{XXX}$ and $B_{\ell_1 \ell_2 \ell_3}^{BBX} \equiv B_{(tts) \ell_1 \ell_2 \ell_3}^{BBX} $. We focus on these cases, since they represent the best representative combinations to be compared.  Here we have assumed a very weak non-Gaussian signal, so that the variance can be expressed with the products of the angular power spectra $C_\ell^{XX}$ and $C_\ell^{BB}$. Furthermore, in order to derive ${\cal F}^{BBX}$, we have used the fact that the expected signal of the cross-correlation $XB$ is undetectably small, i.e., $C_\ell^{XX} C_\ell^{BB} \gg (C_\ell^{XB})^2 $. In the following, we consider a full-sky noiseless cosmic-variance-limited-level (CVL-level) experiment. In this case, $C_\ell^{XX}$ and $C_\ell^{BB}$ are determined by the signal computed from theory (and when specified it includes the contribution from lensing, see later). The expected $1\sigma$ errors on $\Pi$ for $r = 10^{-2}$ and $10^{-3}$, computed according to $\Delta \Pi^{X_1 X_2 X_3} = 1/\sqrt{{\cal F}^{X_1 X_2 X_3}}$, are shown in figure~\ref{fig:error_Pi}.

From the $TTT$ results, we find that the usual scaling relation for the squeezed-type non-Gaussianity case, $\Delta \Pi^{TTT} \propto \ell_{\rm max}^{-1}$ \cite{Komatsu:2001rj,Shiraishi:2010kd,Domenech:2017kno}, stops at $\ell_{\rm max} \sim 100$ corresponding to the end of the large-scale amplification due to the integrated Sachs-Wolfe effect induced by gravitational waves \cite{Pritchard:2004qp}. The same suppression was confirmed in the tensor-tensor-tensor bispectrum case \cite{Shiraishi:2011dh,Shiraishi:2011st,Shiraishi:2012sn,Shiraishi:2013kxa,Namba:2015gja}. At very low $\ell$'s, ${\cal T}_{\ell (t)}^T$ is comparable in size to ${\cal T}_{\ell (s)}^T$, so, e.g., $B_{(tts) \ell_1 \ell_2 \ell_3}^{TTT} / B_{(sss) \ell_1 \ell_2 \ell_3}^{TTT} \sim \Braket{\gamma \gamma \zeta} / \Braket{\zeta \zeta \zeta}$ becomes a good approximation. Considering the comparison with the usual scalar-mode local-type non-Gaussianity case \cite{Komatsu:2001rj}, we therefore have $B_{(tts) \ell_1 \ell_2 \ell_3}^{TTT} / B_{(sss) \ell_1 \ell_2 \ell_3}^{TTT} \sim 10^{-3} \times  (r^2 \Pi) / f_{\rm NL}$ (see, e.g., eq.~(5.62) of ref.~\cite{Bartolo:2017szm}). From this, we find a transforming formula $\Delta \Pi^{TTT} \sim 10^3 r^{-2} \Delta f_{\rm NL}^{TTT}$. Substituting the expected $1\sigma$ error on $f_{\rm NL}$ obtained in the literature, $\Delta f_{\rm NL}^{TTT}(\ell_{\rm max} = 10) \sim 10^3$ \cite{Komatsu:2001rj} into this equation, we can derive $\Delta \Pi^{TTT}(\ell_{\rm max} = 10) \sim 10^6 r^{-2}$. As expected, this fully agrees with the results in figure~\ref{fig:error_Pi}. Note that $\Delta \Pi^{TTT} \propto r^{-2}$ exactly holds because of $\hat{B}_{\ell_1 \ell_2 \ell_3}^{TTT} \propto r^2$. The similar features are also seen in the $EEE$ case.

The Fisher matrix from $BBT$ or $BBE$ is computed assuming two kinds of cases where the B-mode polarization data is fully-delensed or undelensed. In the former case, the errors are expected to scale like $\Delta \Pi^{BBX} \propto r^{-1}$ because of $C_\ell^{BB} \propto r$ and $\hat{B}_{\ell_1 \ell_2 \ell_3}^{BBX} \propto r^2$. Moreover, at very low $\ell$'s, $\hat{B}_{\ell_1 \ell_2 \ell_3}^{BBX} / \hat{B}_{\ell_1 \ell_2 \ell_3}^{XXX} \sim 1$ and $C_\ell^{BB} / C_\ell^{XX} \sim r$ hold because of the small shape difference between ${\cal T}_{\ell (t)}^{T}$, ${\cal T}_{\ell (t)}^{E}$, ${\cal T}_{\ell (t)}^{B}$, ${\cal T}_{\ell (s)}^{T}$ and ${\cal T}_{\ell (s)}^{E}$; thus, $\Delta \Pi^{BBX} / \Delta \Pi^{XXX} \sim r$ is expected to hold. All these scaling are actually confirmed from the solid lines in figure~\ref{fig:error_Pi}, justifying our numerical results for $BBT$ and $BBE$.

The former analysis tells us that if delensing perfectly works, the squeezed-type scaling $\Delta \Pi^{BBX} \propto \ell_{\rm max}^{-1}$ is still maintained beyond $\ell_{\rm max} \sim 100$.%
\footnote{
Very roughly, the effects of the transfer functions in $\hat{B}_{\ell_1 \ell_2 \ell_3}^{BBX}$ and $\sqrt{C_{\ell_1}^{BB} C_{\ell_2}^{BB} C_{\ell_3}^{XX}}$ of eq.~\eqref{eq:fish_BBX} cancel each other out; thus, one can obtain this result following closely the estimate for local bispectra given in refs.~\cite{Komatsu:2001rj,Babich:2004yc}.
}
From the above estimates and our numerical results, we can therefore expect that
\begin{eqnarray}
  \Delta \Pi^{BBT} \sim \Delta \Pi^{BBE} \sim 10^6 \left(\frac{0.01}{r}\right) \left( \frac{500}{\ell_{\rm max}} \right) \, . \label{eq:dPi_BBT}
\end{eqnarray}

Even in the perfectly-delensed situation, such a rapid sensitivity improvement by increasing $\ell_{\rm max}$ is not expected when estimating the usual power spectrum chirality parameter,
  \begin{eqnarray}
    \chi \equiv \frac{\Braket{\gamma^{(+2)} \gamma^{(+2)}} - \Braket{\gamma^{(-2)} \gamma^{(-2)}}}{\Braket{\gamma^{(+2)} \gamma^{(+2)}} + \Braket{\gamma^{(-2)} \gamma^{(-2)}}},
  \end{eqnarray}
  from $C_\ell^{TB}$ and $C_\ell^{EB}$ \cite{Gluscevic:2010vv,Gerbino:2016mqb,Shiraishi:2013kxa}. The Fisher matrix in this case is expressed as
  \begin{eqnarray}
    {\cal F}^{XB} = \sum_{\ell = 2}^{\ell_{\rm max}} (2\ell + 1)
    \frac{\left( \hat{C}_\ell^{XB} \right)^2}{C_\ell^{XX} C_\ell^{BB}},
  \end{eqnarray}
  where $\hat{C}_\ell^{XB} \equiv \partial C_\ell^{XB} / \partial \chi = C_\ell^{XB} / \chi$. Especially for high $\ell$, $\hat{C}_\ell^{XB}$ and $C_\ell^{BB}$ (that are sourced by the tensor modes alone) are subdominant compared with $C_\ell^{XX}$ (that is also generated from the scalar mode); hence, $(\hat{C}_\ell^{XB})^2 / (C_\ell^{XX} C_\ell^{BB})$ is highly suppressed. This actually prevents ${\cal F}^{XB}$ from growing. In contrast, in eq.~\eqref{eq:fish_BBX}, both $\hat{B}_{\ell_1 \ell_2 \ell_3}^{BBX}$ and $\sqrt{C_{\ell_1}^{BB} C_{\ell_2}^{BB} C_{\ell_3}^{XX}}$ come from two tensors and one scalar and therefore $( \hat{B}_{\ell_1 \ell_2 \ell_3}^{BBX} )^2 / (C_{\ell_1}^{BB} C_{\ell_2}^{BB} C_{\ell_3}^{XX})$ does not decay for high $\ell$, enhancing ${\cal F}^{BBX}$. This indicates that CMB bispectra represent a promising observable to test and measure the chirality of gravitational waves arising from parity-violation effects.

In contrast, the presence of the lensing B-mode can degrade the sensitivity, as expected and as shown in figure~\ref{fig:error_Pi}.

There is a further interesting aspect to point out. In real data analysis, one may be concerned about the contaminations due to late-time secondary contributions such as the so-called ISW-lensing and polarization-lensing bispectra \cite{Hu:2000ee,Lewis:2011fk}. Fortunately, however, these effects completely vanish in the multipole domain under consideration \eqref{eq:P-odd_rule} (because these late-time effects are not parity-violating sources). Hence, the extra process of subtracting such secondary contributions is indeed not required.

\section{Conclusions}\label{sec:conclusions}

In this work we studied the observational prospects of measuring parity-breaking bispectra statistics arising from a Chern-Simons gravitational term 
coupled to the inflaton field through a generic coupling function $f(\phi)$. 
Our final Fisher matrix forecasts, eq.~\eqref{eq:dPi_BBT}, tell us that, if $r = 10^{-2}$, by the measurement of $BBT$ or $BBE$ CMB angular bispectra, $\Pi \sim 10^6$ is testable. From the theoretical point of view, such a large chirality in $\Braket{ \gamma \gamma \zeta }$ bispectrum can be realized in the case of a time dependent Chern-Simons mass, since $\Pi$ can be treated as a free parameter. Thus, $BBT$ or $BBE$ CMB angular bispectra could become in the future an essential observable for testing Chern-Simons gravity with a time dependent Chern-Simons mass during inflation. Moreover, we showed that  an improvement in the angular resolution of the experiment could in principle enhance the minimum testable value of $\Pi$. In fact, the $1\sigma$ ideal error on $\Pi$ scales as $\ell_{\rm max}^{-1}$ (contrary to what happens when estimating chirality from power spectra $C_\ell^{TB}$ and $C_\ell^{EB}$). However, realistically speaking the B-modes coming from gravitational lensing degrade the small scale contribution to the S/N ratio in a way that a saturation is achieved for $\ell>10^2$ (see figure~\ref{fig:error_Pi}). We leave further discussions and considerations about the CMB delensing for future work.


\acknowledgments

G.\,O. thanks Ippei Obata for useful discussions. M.\,S. was supported by JSPS Grant-in-Aid for Research Activity Start-up Grant Number 17H07319. Numerical computations by M.\,S. were in part carried out on Cray XC50 at Center for Computational Astrophysics, National Astronomical Observatory of Japan. N.B. acknowledges partial financial support by the ASI/INAF Agreement I/072/09/0 for the Planck LFI Activity of Phase E2. We acknowledge also financial support by ASI Grant 2016-24-H.0.





\bibliography{paper}

\providecommand{\href}[2]{#2}\begingroup\raggedright\begin{thebibliography}{10}

\bibitem{Bartolo:2017szm}
N.~Bartolo and G.~Orlando, \emph{{Parity breaking signatures from a
  Chern-Simons coupling during inflation: the case of non-Gaussian
  gravitational waves}},
  \href{http://dx.doi.org/10.1088/1475-7516/2017/07/034}{\emph{JCAP} {\bf 1707}
  (2017) 034}, [\href{http://arxiv.org/abs/1706.04627}{{\tt 1706.04627}}].

\bibitem{Weinberg:2008}
S.~{Weinberg}, \emph{{Effective field theory for inflation}},
  \href{http://dx.doi.org/10.1103/PhysRevD.77.123541}{\emph{\prd} {\bf 77}
  (June, 2008) 123541}, [\href{http://arxiv.org/abs/0804.4291}{{\tt
  0804.4291}}].

\bibitem{Maldacena:2011}
J.~M. {Maldacena} and G.~L. {Pimentel}, \emph{{On graviton non-gaussianities
  during inflation}},
  \href{http://dx.doi.org/10.1007/JHEP09(2011)045}{\emph{Journal of High Energy
  Physics} {\bf 2011} (Sept., 2011) 45},
  [\href{http://arxiv.org/abs/1104.2846}{{\tt 1104.2846}}].

\bibitem{Shiraishi:2011st}
M.~Shiraishi, D.~Nitta and S.~Yokoyama, \emph{{Parity Violation of Gravitons in
  the CMB Bispectrum}},
  \href{http://dx.doi.org/10.1143/PTP.126.937}{\emph{Prog.Theor.Phys.} {\bf
  126} (2011) 937--959}, [\href{http://arxiv.org/abs/1108.0175}{{\tt
  1108.0175}}].

\bibitem{Soda:2011}
J.~{Soda}, H.~{Kodama} and M.~{Nozawa}, \emph{{Parity violation in graviton
  non-gaussianity}},
  \href{http://dx.doi.org/10.1007/JHEP08(2011)067}{\emph{Journal of High Energy
  Physics} {\bf 2011} (Aug., 2011) 67},
  [\href{http://arxiv.org/abs/1106.3228}{{\tt 1106.3228}}].

\bibitem{Lue:1999}
A.~{Lue}, L.~{Wang} and M.~{Kamionkowski}, \emph{{Cosmological Signature of New
  Parity-Violating Interactions}},
  \href{http://dx.doi.org/10.1103/PhysRevLett.83.1506}{\emph{Physical Review
  Letters} {\bf 83} (Aug., 1999) 1506--1509},
  [\href{http://arxiv.org/abs/astro-ph/9812088}{{\tt astro-ph/9812088}}].

\bibitem{Jackiw:2003}
R.~{Jackiw} and S.-Y. {Pi}, \emph{{Chern-Simons modification of general
  relativity}}, \href{http://dx.doi.org/10.1103/PhysRevD.68.104012}{\emph{\prd}
  {\bf 68} (Nov., 2003) 104012},
  [\href{http://arxiv.org/abs/gr-qc/0308071}{{\tt gr-qc/0308071}}].

\bibitem{Alexander:2005}
S.~{Alexander} and J.~{Martin}, \emph{{Birefringent gravitational waves and the
  consistency check of inflation}},
  \href{http://dx.doi.org/10.1103/PhysRevD.71.063526}{\emph{\prd} {\bf 71}
  (Mar., 2005) 063526}, [\href{http://arxiv.org/abs/hep-th/0410230}{{\tt
  hep-th/0410230}}].

\bibitem{Alexander:2006}
S.~H. {Alexander}, M.~E. {Peskin} and M.~M. {Sheikh-Jabbari},
  \emph{{Leptogenesis from Gravity Waves in Models of Inflation}},
  \href{http://dx.doi.org/10.1103/PhysRevLett.96.081301}{\emph{Physical Review
  Letters} {\bf 96} (Feb., 2006) 081301},
  [\href{http://arxiv.org/abs/hep-th/0403069}{{\tt hep-th/0403069}}].

\bibitem{Alexander:2007}
S.~H. {Alexander}, M.~{Peskin} and M.~M. {Sheikh-Jabbari},
  \emph{{Gravi-Leptogenesis: Leptogenesis from Gravity Waves in Pseudo-scalar
  Driven Inflation Models}}, {\emph{ArXiv High Energy Physics - Phenomenology
  e-prints} (Jan., 2007) }, [\href{http://arxiv.org/abs/hep-ph/0701139}{{\tt
  hep-ph/0701139}}].

\bibitem{Satoh:2008}
M.~{Satoh} and J.~{Soda}, \emph{{Higher curvature corrections to primordial
  fluctuations in slow-roll inflation}},
  \href{http://dx.doi.org/10.1088/1475-7516/2008/09/019}{\emph{\jcap} {\bf 9}
  (Sept., 2008) 019}, [\href{http://arxiv.org/abs/0806.4594}{{\tt 0806.4594}}].

\bibitem{Alexander:2009}
S.~{Alexander} and N.~{Yunes}, \emph{{Chern-Simons modified general
  relativity}},
  \href{http://dx.doi.org/10.1016/j.physrep.2009.07.002}{\emph{\physrep} {\bf
  480} (Aug., 2009) 1--55}, [\href{http://arxiv.org/abs/0907.2562}{{\tt
  0907.2562}}].

\bibitem{Satoh:2010}
M.~{Satoh}, \emph{{Slow-roll inflation with the Gauss-Bonnet and Chern-Simons
  corrections}},
  \href{http://dx.doi.org/10.1088/1475-7516/2010/11/024}{\emph{\jcap} {\bf 11}
  (Nov., 2010) 024}, [\href{http://arxiv.org/abs/1008.2724}{{\tt 1008.2724}}].

\bibitem{Malenknejad:2012}
A.~{Maleknejad}, M.~{Noorbala} and M.~M. {Sheikh-Jabbari}, \emph{{Leptogenesis
  in Inflationary models with Non-Abelian Gauge Fields}}, {\emph{ArXiv
  e-prints} (Aug., 2012) }, [\href{http://arxiv.org/abs/1208.2807}{{\tt
  1208.2807}}].

\bibitem{Myung:2014}
Y.~S. {Myung} and T.~{Moon}, \emph{{Primordial massive gravitational waves from
  Einstein-Chern-Simons-Weyl gravity}},
  \href{http://dx.doi.org/10.1088/1475-7516/2014/08/061}{\emph{\jcap} {\bf 8}
  (Aug., 2014) 061}, [\href{http://arxiv.org/abs/1406.4367}{{\tt 1406.4367}}].

\bibitem{Alexander:2016}
S.~{Alexander}, \emph{{Inflationary birefringence and baryogenesis}},
  \href{http://dx.doi.org/10.1142/S0218271816400137}{\emph{International
  Journal of Modern Physics D} {\bf 25} (June, 2016) 1640013},
  [\href{http://arxiv.org/abs/1604.00703}{{\tt 1604.00703}}].

\bibitem{Baumann:2016}
D.~{Baumann}, H.~{Lee} and G.~L. {Pimentel}, \emph{{High-scale inflation and
  the tensor tilt}},
  \href{http://dx.doi.org/10.1007/JHEP01(2016)101}{\emph{Journal of High Energy
  Physics} {\bf 2016} (Jan., 2016) 101},
  [\href{http://arxiv.org/abs/1507.07250}{{\tt 1507.07250}}].

\bibitem{Kawai:2017}
S.~{Kawai} and J.~{Kim}, \emph{{Gauss-Bonnet Chern-Simons gravitational wave
  leptogenesis}}, {\emph{ArXiv e-prints} (Feb., 2017) },
  [\href{http://arxiv.org/abs/1702.07689}{{\tt 1702.07689}}].

\bibitem{Cai:2017}
Y.~{Cai}, Y.-T. {Wang} and Y.-S. {Piao}, \emph{{Chirality oscillation of
  primordial gravitational waves during inflation}},
  \href{http://dx.doi.org/10.1007/JHEP03(2017)024}{\emph{Journal of High Energy
  Physics} {\bf 3} (Mar., 2017) 24},
  [\href{http://arxiv.org/abs/1608.06508}{{\tt 1608.06508}}].

\bibitem{Afkhami-Jeddi:2018own}
N.~Afkhami-Jeddi, S.~Kundu and A.~Tajdini, \emph{{A Conformal Collider for
  Holographic CFTs}},  \href{http://arxiv.org/abs/1805.07393}{{\tt
  1805.07393}}.

\bibitem{Gerbino:2016mqb}
M.~Gerbino, A.~Gruppuso, P.~Natoli, M.~Shiraishi and A.~Melchiorri,
  \emph{{Testing chirality of primordial gravitational waves with Planck and
  future CMB data: no hope from angular power spectra}},
  \href{http://dx.doi.org/10.1088/1475-7516/2016/07/044}{\emph{JCAP} {\bf 1607}
  (2016) 044}, [\href{http://arxiv.org/abs/1605.09357}{{\tt 1605.09357}}].

\bibitem{Shiraishi:2012sn}
M.~Shiraishi, \emph{{Parity violation of primordial magnetic fields in the CMB
  bispectrum}},
  \href{http://dx.doi.org/10.1088/1475-7516/2012/06/015}{\emph{JCAP} {\bf 1206}
  (2012) 015}, [\href{http://arxiv.org/abs/1202.2847}{{\tt 1202.2847}}].

\bibitem{Gluscevic:2010vv}
V.~Gluscevic and M.~Kamionkowski, \emph{{Testing Parity-Violating Mechanisms
  with Cosmic Microwave Background Experiments}},
  \href{http://dx.doi.org/10.1103/PhysRevD.81.123529}{\emph{Phys. Rev.} {\bf
  D81} (2010) 123529}, [\href{http://arxiv.org/abs/1002.1308}{{\tt
  1002.1308}}].

\bibitem{Kamionkowski:2010rb}
M.~Kamionkowski and T.~Souradeep, \emph{{The Odd-Parity CMB Bispectrum}},
  \href{http://dx.doi.org/10.1103/PhysRevD.83.027301}{\emph{Phys.Rev.} {\bf
  D83} (2011) 027301}, [\href{http://arxiv.org/abs/1010.4304}{{\tt
  1010.4304}}].

\bibitem{Shiraishi:2013kxa}
M.~Shiraishi, A.~Ricciardone and S.~Saga, \emph{{Parity violation in the CMB
  bispectrum by a rolling pseudoscalar}},
  \href{http://dx.doi.org/10.1088/1475-7516/2013/11/051}{\emph{JCAP} {\bf 1311}
  (2013) 051}, [\href{http://arxiv.org/abs/1308.6769}{{\tt 1308.6769}}].

\bibitem{Namba:2015gja}
R.~Namba, M.~Peloso, M.~Shiraishi, L.~Sorbo and C.~Unal, \emph{{Scale-dependent
  gravitational waves from a rolling axion}},
  \href{http://dx.doi.org/10.1088/1475-7516/2016/01/041}{\emph{JCAP} {\bf 1601}
  (2016) 041}, [\href{http://arxiv.org/abs/1509.07521}{{\tt 1509.07521}}].

\bibitem{Shiraishi:2016yun}
M.~Shiraishi, C.~Hikage, R.~Namba, T.~Namikawa and M.~Hazumi, \emph{{Testing
  statistics of the CMB B -mode polarization toward unambiguously establishing
  quantum fluctuation of the vacuum}},
  \href{http://dx.doi.org/10.1103/PhysRevD.94.043506}{\emph{Phys. Rev.} {\bf
  D94} (2016) 043506}, [\href{http://arxiv.org/abs/1606.06082}{{\tt
  1606.06082}}].

\bibitem{Shiraishi:2016mok}
M.~Shiraishi, \emph{{Parity violation in the CMB trispectrum from the scalar
  sector}}, \href{http://dx.doi.org/10.1103/PhysRevD.94.083503}{\emph{Phys.
  Rev.} {\bf D94} (2016) 083503}, [\href{http://arxiv.org/abs/1608.00368}{{\tt
  1608.00368}}].

\bibitem{Saito:2007kt}
S.~Saito, K.~Ichiki and A.~Taruya, \emph{{Probing polarization states of
  primordial gravitational waves with CMB anisotropies}},
  \href{http://dx.doi.org/10.1088/1475-7516/2007/09/002}{\emph{JCAP} {\bf 0709}
  (2007) 002}, [\href{http://arxiv.org/abs/0705.3701}{{\tt 0705.3701}}].

\bibitem{Shiraishi:2014ila}
M.~Shiraishi, M.~Liguori and J.~R. Fergusson, \emph{{Observed parity-odd CMB
  temperature bispectrum}},
  \href{http://dx.doi.org/10.1088/1475-7516/2015/01/007}{\emph{JCAP} {\bf 1501}
  (2015) 007}, [\href{http://arxiv.org/abs/1409.0265}{{\tt 1409.0265}}].

\bibitem{Ade:2015ava}
{\scshape Planck} collaboration, P.~A.~R. Ade et~al., \emph{{Planck 2015
  results. XVII. Constraints on primordial non-Gaussianity}},
  \href{http://dx.doi.org/10.1051/0004-6361/201525836}{\emph{Astron.
  Astrophys.} {\bf 594} (2016) A17},
  [\href{http://arxiv.org/abs/1502.01592}{{\tt 1502.01592}}].

\bibitem{Bartolo:2016ami}
N.~Bartolo et~al., \emph{{Science with the space-based interferometer LISA. IV:
  Probing inflation with gravitational waves}},
  \href{http://dx.doi.org/10.1088/1475-7516/2016/12/026}{\emph{JCAP} {\bf 1612}
  (2016) 026}, [\href{http://arxiv.org/abs/1610.06481}{{\tt 1610.06481}}].

\bibitem{Weinberg:2005}
S.~{Weinberg}, \emph{{Quantum contributions to cosmological correlations}},
  \href{http://dx.doi.org/10.1103/PhysRevD.72.043514}{\emph{\prd} {\bf 72}
  (Aug., 2005) 043514}, [\href{http://arxiv.org/abs/hep-th/0506236}{{\tt
  hep-th/0506236}}].

\bibitem{Maldacena:2003}
J.~{Maldacena}, \emph{{Non-gaussian features of primordial fluctuations in
  single field inflationary models}},
  \href{http://dx.doi.org/10.1088/1126-6708/2003/05/013}{\emph{Journal of High
  Energy Physics} {\bf 2003} (May, 2003) 013},
  [\href{http://arxiv.org/abs/astro-ph/0210603}{{\tt astro-ph/0210603}}].

\bibitem{Collins:2011}
H.~{Collins}, \emph{{Primordial non-Gaussianities from inflation}},
  {\emph{arXiv e-prints} (Jan., 2011) arXiv:1101.1308},
  [\href{http://arxiv.org/abs/1101.1308}{{\tt 1101.1308}}].

\bibitem{Chen:2007}
X.~{Chen}, M.-x. {Huang}, S.~{Kachru} and G.~{Shiu}, \emph{{Observational
  signatures and non-Gaussianities of general single-field inflation}},
  \href{http://dx.doi.org/10.1088/1475-7516/2007/01/002}{\emph{Journal of
  Cosmology and Astro-Particle Physics} {\bf 2007} (Jan., 2007) 002},
  [\href{http://arxiv.org/abs/hep-th/0605045}{{\tt hep-th/0605045}}].

\bibitem{Shiraishi:2010sm}
M.~Shiraishi, S.~Yokoyama, D.~Nitta, K.~Ichiki and K.~Takahashi,
  \emph{{Analytic formulae of the CMB bispectra generated from non-Gaussianity
  in the tensor and vector perturbations}},
  \href{http://dx.doi.org/10.1103/PhysRevD.82.103505}{\emph{Phys. Rev.} {\bf
  D82} (2010) 103505}, [\href{http://arxiv.org/abs/1003.2096}{{\tt
  1003.2096}}].

\bibitem{Shiraishi:2010kd}
M.~Shiraishi, D.~Nitta, S.~Yokoyama, K.~Ichiki and K.~Takahashi, \emph{{CMB
  Bispectrum from Primordial Scalar, Vector and Tensor non-Gaussianities}},
  \href{http://dx.doi.org/10.1143/PTP.125.795}{\emph{Prog. Theor. Phys.} {\bf
  125} (2011) 795--813}, [\href{http://arxiv.org/abs/1012.1079}{{\tt
  1012.1079}}].

\bibitem{Shiraishi:2016ads}
M.~Shiraishi, \emph{{Search for primordial symmetry breakings in CMB}},
  \href{http://dx.doi.org/10.1142/S0217732316400034}{\emph{Mod. Phys. Lett.}
  {\bf A31} (2016) 1640003}.

\bibitem{Komatsu:2001rj}
E.~Komatsu and D.~N. Spergel, \emph{{Acoustic signatures in the primary
  microwave background bispectrum}},
  \href{http://dx.doi.org/10.1103/PhysRevD.63.063002}{\emph{Phys. Rev.} {\bf
  D63} (2001) 063002}, [\href{http://arxiv.org/abs/astro-ph/0005036}{{\tt
  astro-ph/0005036}}].

\bibitem{Domenech:2017kno}
G.~Dom\`enech, T.~Hiramatsu, C.~Lin, M.~Sasaki, M.~Shiraishi and Y.~Wang,
  \emph{{CMB Scale Dependent Non-Gaussianity from Massive Gravity during
  Inflation}},
  \href{http://dx.doi.org/10.1088/1475-7516/2017/05/034}{\emph{JCAP} {\bf 1705}
  (2017) 034}, [\href{http://arxiv.org/abs/1701.05554}{{\tt 1701.05554}}].

\bibitem{Pritchard:2004qp}
J.~R. Pritchard and M.~Kamionkowski, \emph{{Cosmic microwave background
  fluctuations from gravitational waves: An Analytic approach}},
  \href{http://dx.doi.org/10.1016/j.aop.2005.03.005}{\emph{Annals Phys.} {\bf
  318} (2005) 2--36}, [\href{http://arxiv.org/abs/astro-ph/0412581}{{\tt
  astro-ph/0412581}}].

\bibitem{Shiraishi:2011dh}
M.~Shiraishi, D.~Nitta, S.~Yokoyama, K.~Ichiki and K.~Takahashi, \emph{{Cosmic
  microwave background bispectrum of tensor passive modes induced from
  primordial magnetic fields}},
  \href{http://dx.doi.org/10.1103/PhysRevD.83.123003}{\emph{Phys. Rev.} {\bf
  D83} (2011) 123003}, [\href{http://arxiv.org/abs/1103.4103}{{\tt
  1103.4103}}].

\bibitem{Babich:2004yc}
D.~Babich and M.~Zaldarriaga, \emph{{Primordial bispectrum information from CMB
  polarization}},
  \href{http://dx.doi.org/10.1103/PhysRevD.70.083005}{\emph{Phys. Rev.} {\bf
  D70} (2004) 083005}, [\href{http://arxiv.org/abs/astro-ph/0408455}{{\tt
  astro-ph/0408455}}].

\bibitem{Hu:2000ee}
W.~Hu, \emph{{Weak lensing of the CMB: A harmonic approach}},
  \href{http://dx.doi.org/10.1103/PhysRevD.62.043007}{\emph{Phys. Rev.} {\bf
  D62} (2000) 043007}, [\href{http://arxiv.org/abs/astro-ph/0001303}{{\tt
  astro-ph/0001303}}].

\bibitem{Lewis:2011fk}
A.~Lewis, A.~Challinor and D.~Hanson, \emph{{The shape of the CMB lensing
  bispectrum}},
  \href{http://dx.doi.org/10.1088/1475-7516/2011/03/018}{\emph{JCAP} {\bf 1103}
  (2011) 018}, [\href{http://arxiv.org/abs/1101.2234}{{\tt 1101.2234}}].

\end{thebibliography}\endgroup
\end{document}